\documentclass[epj, onecolumn]{svjour}
\usepackage{graphics}
\usepackage{amsmath}
\usepackage{graphicx}
\usepackage{dcolumn}
\usepackage{bm}
\usepackage{amssymb}

\begin{document}

\title{Quasinormal Modes of Lovelock Black Holes}

\author{C.B. PRASOBH\thanks{e-mail: prasobhcb@outlook.com}\and V.C. KURIAKOSE\thanks{e-mail: vck@cusat.ac.in}% etc
}

\offprints{}          % Insert a name or remove this line
\institute{Department of Physics, Cochin University of Science and Technology, Cochin 682202, India}
\date{Received: date / Revised version: date}
\abstract{
The quasinormal modes of metric perturbations in asymptotically flat black hole 
spacetimes in the Lovelock model are 
calculated for different spacetime dimensions and higher orders of curvature. 
It is analytically established that in the asymptotic limit $l \rightarrow \infty$, the imaginary parts of 
the quasi normal frequencies become constant for tensor, scalar as well as vector perturbations. Numerical 
calculation shows that this indeed is the case. Also, the real and imaginary parts of the quasinormal modes 
are seen to increase as the order of the theory $k$ increases.
The real part of the modes decreases as the spacetime dimension $d$ increases, indicating the presence of lower
frequency modes in higher dimensions. Also, it is seen that the modes are roughly isospectral at very high 
values of the spacetime dimension $d$.%
\PACS{
      {04.50.Gh}{ Higher-dimensional black holes, black strings, and related objects}   \and
      {04.50.-h}{Higher-dimensional gravity and other theories of gravity}\and
      {04.50.Kd}{Modified theories of gravity}
     } % end of PACS codes
} %end of abstract
\maketitle

\section{Introduction}
\label{intro}
\paragraph{}

Quasinormal modes (QNMs) are damped oscillatory modes of a field that perturbs the spacetime 
metric in the vicinity of a black hole. 
They depend only on the parameters of the black hole, and not on the nature of the 
perturbing field. This makes them ideal tools 
to study the physics of black holes, which are otherwise impossible to observe by their 
very definition. The long-lived modes in asymptotically flat spacetimes surrounding black 
holes are expected to be observed in future by gravitational wave detectors. Different models of 
gravity predict different ``quasinormal signatures'' of their respective spacetimes and 
the experimental observation of these modes may well put to rest the problem of selecting 
the most suitable model for gravity from existing (numerous) ones.\\

The research on QNMs is decades old with an extensive literature
(for example, \cite{berti,nollert,kokkotas,konoplya1,konoplya2,cho,mashhoon1,mashhoon2,schutz_will,iyer_will,leaver} and 
references therein). 
The quasinormal behavior in first order theories of gravity such as 
the General Theory of Relativity (GTR) is particularly well studied with its asymptotic 
behavior firmly established both numerically and analytically 
\cite{motl}. The asymptotic quasinormal modes of perturbations in GTR have their real parts approach a constant value, while 
the imaginary parts increase indefinitely.  
These modes are significant from the standpoint of quantum theories of gravity since they help us to compute the area 
spectrum 
and subsequently the entropy of the black hole event horizons, which, in GTR, are known to be equally spaced. The asymptotic 
behavior of the modes, observed numerically, can help one analytically determine the precise form of these modes in terms 
of the parameters of the theory later. This has been demonstrated in \cite{motl}, where the decision to compute the monodromy 
along the Stokes line was made because of the asymptotic behavior mentioned above. Thus it would be highly interesting to see 
how the quasinormal modes behave asymptotically in any model of gravity that one considers.\\

The connection between geodesic stability and quasinormal modes in black hole spacetimes has been known for a long time 
(\cite{cardoso_miranda_berti,kokkotas,nollert,ferrari_mashhoon,berti_kokkotas} among others). 
These studies reveal the connection between quasinormal modes of black hole spacetimes and the dynamics of null particles 
in an unstable circular orbit around the black hole, with its energy slowly leaking out. The relation is most clearly established 
in \cite{cardoso_miranda_berti} for any static, spherically symmetric and asymptotically flat spacetime, according to which 
the quasinormal frequencies $\omega_{asy}$ in the asymptotic limit $(l \rightarrow \infty )$ is given by

\begin{equation}
 \omega_{asy}=\Omega_c l - i (n + \frac{1}{2} ) | \lambda |,
\end{equation}
 
where $\Omega_c$ and $\lambda$ are the angular velocity at the unstable null geodesic and the principal Lyapunov exponent 
which is related to the time scale of energy decay in the orbit. \\

The actual number of spacetime dimensions is predicted to be higher than four by string theory and it has led to 
attempts at 
developing models of gravity in higher dimensions. In these higher dimensional spacetimes, GTR no longer is the most 
general model of gravity. Generalizations of GTR are naturally attempted by adding higher order curvature correction 
terms to the 
Einstein-Hilbert action. Among such generalizations to GTR, the Lovelock model \cite{lovelock,wheeler}, 
considered as a natural generalization of the GTR to higher dimensions and orders of 
curvature, is particularly interesting since it yields field equations of second order 
that are free of ghosts. The Lovelock Lagrangian consists of dimensionally continued curvature terms of orders one and above. 
The resulting theories are 
labeled by the order of the maximum-ordered term, $k$, which in turn is determined by 
the dimension of the spacetime $d$, by $k=[\frac{d-1}{2}]$ where $[x]$ denotes the integer 
part of $x$. Black hole solutions to the theory in general contain many branches that depend on the values 
of the higher order coupling constants \cite{takahashimaster}. It is known \cite{takahashiflatunstable} that 
the metric perturbations to the most general, asymptotically 
flat Lovelock spacetime are unstable in the ultraviolet region. Therefore it is necessary to impose further 
constraints to select a suitable set of Lovelock theories which would permit stable perturbations. 
Such maximally symmetric, asymptotically flat as well as AdS spacetimes have been known for a long time \cite{bhscan}. 
 \\

In this work, we compute the quasinormal modes of metric perturbations to the metric of such maximally 
symmetric spacetimes using 
the sixth order WKB method \cite{konoplya_order6}. We analytically determine the asymptotic form of these modes using the 
above-mentioned null geodesic method. The paper is organized as follows: in Sect. \ref{null}, we describe the essential 
details of the null geodesic method used to compute the asymptotic form of the modes. In Sect. \ref{lltheory}, we describe the 
class of Lovelock theories for which the modes are computed and the WKB expression of numerical computation. The relation 
between the asymptotic quasinormal modes and the null geodesic parameters is expressed in Sect. \ref{qnm_from_null}. 
The results of the calculation are discussed in Sect. \ref{discussion}. We summarize the main results of the work in Sect. \ref{conclusion}.

\section{Geodesic stability}\label{null}

Consider the general stationary and spherically symmetric metric

\begin{equation}\label{general_metric}
 ds^{2}= f(r)dt^{2}- \frac{1}{g(r)} dr^{2}-r^{2}d\Omega_{d-2}^2,
\end{equation}

where $f(r)$ and $g(r)$ are solutions of the Lovelock field equations \cite{takahashimaster}. $d\Omega_{d-2}^2$ represents the metric of the 
spherically symmetric background. For this metric, we have the Lagrangian in the form,\cite{chandrasekhar}

\begin{equation}
 2{\mathcal L}=f(r)\,\dot{t}^2-\frac{1}{g(r)}\dot{r}^2-r^2\dot{\varphi}^2,
\end{equation}

where a dot represents derivative with respect to proper time and $\varphi$ is an angular coordinate. For this system, the 
coordinate angular velocity $\Omega_c$ and the principal Lyapunov exponent $\lambda$ for circular null geodesics take the 
form, \cite{cardoso_miranda_berti}

\begin{equation}\label{omega_c}
 \Omega_c=\frac{\dot{\varphi}}{\dot{t}}=\left(\frac{f'_c}{2r_c}\right)^{1/2},
\end{equation}

\begin{equation}\label{lyapunov}
 \lambda= \frac{1}{\sqrt{2}}\sqrt{-\frac{r_c^2}{f_c}\left(\frac{d^2}{dr_*^2}\frac{f}{r^2}\right)_{r=r_c}},
\end{equation}

where the subscript $c$ means that the evaluation is done at the critical radius, $r=r_c$, which satisfies the relation 
$2f-rf'=0$. $r_c$ can be viewed as the innermost circular timelike geodesic, since circular timelike geodesics satisfy 
$2f-rf'>0$. $r_*$ is the tortoise coordinate which satisfies the relation $dr_*=\frac{dr}{\sqrt{g(r)f(r)}}$.

\subsection{The Equations of Perturbation and the WKB method}\label{lltheory}

The action for the class of Lovelock theories, a subset of which are studied in this work, is written in terms of the 
Riemann curvature $R^{ab}=d\omega ^{ab}+\omega _{c}^{a}\omega ^{cb}$ and the vielbein $e^{a}$ as \cite{bhscan,deser,wheeler}

\begin{equation}\label{llaction}
I_{G}=\kappa \int \sum_{p=0}^{k}\alpha _{p}L^{(p)},
\end{equation}
where $\alpha _{p}$ are positive coupling constants and $L^{(p)}$, given by

\begin{equation}\label{Lovelock_l_p}
L^{(p)}=\epsilon _{a_{1}\cdots a_{d}}R^{a_{1}a_{2}}\!\cdot \!\cdot \!\cdot
\!R^{a_{2p-1}a_{2p}}e^{a_{2p+1}}\!\cdot \!\cdot \!\cdot \!e^{a_{d}},
\end{equation}

are the $p^{\textrm{th}}$ order dimensionally continued terms in the Lagrangian, $\epsilon _{a_{1}\cdots a_{d}}$ 
being the Levi-Civita symbol. $\kappa$ is a parameter 
related to the gravitational constant $G_k$ by $\kappa =\frac{1}{2(d-2)!\Omega _{d-2}G_{k}}$, $\Omega _{d-2}$ being the 
volume of the $(d-2)$ dimensional spherically symmetric tangent space with unit curvature.\\ 

The resulting field equations are of the form

\begin{eqnarray}
\epsilon _{ba_{1}\cdots a_{d-1}}\bar{R}^{a_{1}a_{2}}\!\cdot \!\cdot \!\cdot
\!\bar{R}^{a_{2k-1}a_{2k}}e^{a_{2k+1}}\!\cdot \!\cdot \!\cdot \!e^{a_{d-1}}
&=&0  \label{ssymmfieldeq} \\
\epsilon _{aba_{3}\cdots a_{d}}\bar{R}^{a_{3}a_{4}}\!\cdot \!\cdot \!\cdot \!%
\bar{R}^{a_{2k-1}a_{2k}}T^{a_{2k+1}}e^{a_{2k+2}}\!\cdot \!\cdot \!\cdot
\!e^{a_{d-1}} &=&0  \label{ssymmtorsioneq}
\end{eqnarray}

Here, $\bar{R}^{ab}:=R^{ab}+\frac{1}{R^{2}}e^{a}e^{b}$.\\

The quasinormal behavior in a similar class of asymptotically AdS Lovelock theories possessing a unique cosmological 
constant has recently been studied \cite{vck}. It is known \cite{takahashiflatunstable} that the theories in which
 all the 
higher order coupling 
constants $\alpha _{p}$ are positive permit asymptotically flat spacetime solutions 
that suffer from dynamical instability 
against metric perturbations. 
In the present work, we consider a special case. We consider the class of 
theories 
with  $\alpha _{p}$ given by 

\begin{equation}   \label{alphas}
\alpha _{p}=\frac{1}{d-2k}\delta^{k}_{p}
\end{equation}

The static and spherically symmetric black hole solutions of the theory, written in Schwarzschild-like coordinates, 
take the form

\begin{equation}\label{metric}
ds^{2}=f(r)dt^{2}+\frac{dr^{2}}{f(r)}+r^{2}d\Omega
_{d-2}^{2},
\end{equation}

where $f(r)$ is given by

\begin{equation}\label{f}
f(r) =1- \left( \frac{2G_kM}{r^{d-2k-1}}\right) ^{1/k},
\end{equation}

$M$ being the mass of the black hole. It is to be noted that only the cases in which $d-2k-1 \neq 0$ yield black hole 
solutions \cite{bhscan} with their event horizons $r_h$ located at $(2G_kM)^{\frac{1}{d-2k-1}}$. It is 
noted that for the case of $d=4$ and $k=1$, we get the Schwarzschild geometry of GTR. We can therefore consider these 
spacetimes 
as natural generalizations of the former to the case of higher order theories in higher dimensions.\\
 
The master equations obeyed by the metric perturbations for the general Lovelock theory 
were derived in \cite{takahashimaster}. 

The master equation satisfied by the tensor metric perturbation $\delta g_{ij}=r^{2}\phi(t,r)h_{ij}(x^{i})$, after separating the variables 
$\phi (r,t)=\chi (r)e^{-i\omega t}$, takes the form \cite{takahashimaster}

\begin{equation}\label{rteqn}
-f^{2}\chi ^{^{\prime \prime }}-\left( f^{2}\frac{T^{^{\prime \prime }}}{T^{^{\prime }}}+\frac{2f^{2}}{r}+ff^{^{\prime }}\right) \chi ^{^{\prime }}+\frac{(2\kappa +\gamma _{t})f}{(n-2)r}\frac{T^{^{\prime \prime }}}{T^{^{\prime }}}\chi =\omega ^{2}\chi~,
\end{equation}

where the function $T(r)$, for the most general class of Lovelock theories given by (\ref{llaction}) 
with all the constants $\alpha_p$ being positive, is given by the expression

\begin{eqnarray}\label{tdef}
T(r)\equiv r^{n-1}\partial _{\psi }W[\psi ]=r^{n-1} \times \nonumber \\
\left( 1+\displaystyle\sum_{m=2}^{k}\left[ a_{m}\left\{ \prod_{p=1}^{2m-2}(n-p)\right\} \psi^{m-1}\right]\right)~.\label{tdef}
\end{eqnarray}

We write $\Psi (r)=\chi (r)r\sqrt{T^{^{\prime }}(r)}$ and define the tortoise coordinate $r^{\ast }$ 
by $dr^{\ast }=dr/f(r)$ to transform (\ref{rteqn}) to the form

 \begin{equation}\label{schrodinger}
\frac{d^{2}\Psi }{dr^{\ast 2}}+(\Omega ^{2}-V(r))\Psi =0,
\end{equation}

Here, $V(r)=V_t(r)$, the effective potential for tensor perturbations. The tortoise coordinate $r^{\ast }$ is defined by $dr^{\ast }=dr/f(r)$. 
Similar expressions for the vector and scalar type perturbations can be derived easily. The effective potentials $V(r)$ for 
tensor ($V_t(r)$), vector 
($V_v(r)$) and scalar ($V_s(r)$) perturbations are given below:

\begin{equation}
V(r)=\left\{
\begin{array}{ll}
V_{t}(r)=\frac{(2\kappa +\gamma _{t})f}{(n-2)r}\frac{d\ln {T^{^{\prime }}}}{dr}+\frac{1}{r\sqrt{T^{^{\prime }}}}f\frac{d}{dr}\left(f\frac{d}{dr}r\sqrt{T^{^{\prime }}}\right) \\ 
\\
V_v(r)=r\sqrt{T^{'}}f\partial_r \left(f\partial_r\frac{1}{r\sqrt{T^{'}}}\right)+\frac{f}{r}\left(\frac{\gamma_v}{n-1}-\kappa\right)\frac{T^{'}}{T} \\ 
\\
V_s(r)=2\gamma_sf\frac{(rNT)^{'}}{nr^2NT}-f\left(\frac{1}{N}\partial_r(f\partial_rN)+\frac{1}{T}\partial_r(f\partial_rT)\right)\\
\\
+2f^2\left(\frac{N^{'2}}{N^{2}}+\frac{T^{'2}}{T^2}+\frac{N^{'}T^{'}}{NT}\right)\label{effective_potentials}
\end{array}
\right. 
\end{equation} 

Here, $\gamma _{t}=l(l+d-3)-2$, $\gamma _{v}=l(l+d-3)-1$ and $\gamma _{s}=l(l+d-3)$ are the eigenvalues for the tensor, 
vector and scalar harmonics respectively. The functions $T(r)$ and $N(r)$, for the class of theories given by (\ref{alphas}),
 are given by

\begin{eqnarray}\label{aux_functions}
T(r)=\left(\prod_{p=1}^{2k-2}(d-p-2)\right)\left(\frac{2G_kM}{r^{d-1}}\right)^{1-\frac{1}{k}} ,\nonumber\\
N(r)=\frac{2\gamma_s-2(d-2)f+(d-2)rf'}{r\sqrt{T'}}.
\end{eqnarray}
  
\begin{figure}[h]
\centering
\includegraphics[width=0.65\columnwidth]{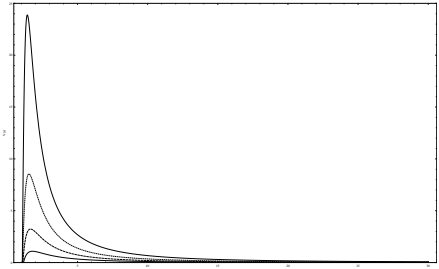}
\caption{Effective potential $V(r)$ vs $r$ for different $k$, from $k=2$ (top) to $k=5$ (bottom), with $d=17$ and $l=7$.} \label{vvsr}
\end{figure} 

Fig. \ref{vvsr} represents the typical variation of the effective potential $V(r)$ outside the event horizon for all 
types of perturbations. 
The different plots are drawn for different values of $k$ which is the tunable parameter for the set of theories 
studied in this work. 
It is noted that the potential is barrier-like for all values of $k$. The height of the barrier is seen to be a decreasing function of 
the order parameter $k$.\\

We now apply the WKB method in order to compute the QNMs of the metric perturbations that obey (\ref{schrodinger}). 
The third order WKB formula for QNMs was derived by Iyer and Will \cite{iyer_will} and was extended to the sixth order 
by 
Konoplya \cite{konoplya_order6}. We use the sixth order formula derived in \cite{konoplya_order6} since it gives 
better accuracy for lower modes.\\

The sixth order formula for computing the QNM $\Omega$ for perturbations obeying (\ref{schrodinger}) is given by

\begin{equation}\label{6th_order_formula}
 \frac{Q_0}{\sqrt{2Q_0^{''}}}-\Lambda_2-\Lambda_3-\Lambda_4-\Lambda_5-\Lambda_6=i\left(n+\frac{1}{2}\right),
\end{equation}

where $n$ is the overtone number and we have used the notation $Q(x)=\Omega^2-V(x)$. 
$Q_0=Q(x_0)$, where $x_0$ is the tortoise coordinate at which the potential attains its peak. Also, prime $(')$ represents 
differentiation with respect to the tortoise coordinate $x$.
The expressions for the correction terms $\Lambda_2, \Lambda_3, \Lambda_4, \Lambda_5$ and $\Lambda_6$ are given in \cite{konoplya_order6} 
and \cite{konoplya_public_URL}. 

\subsection{Asymptotic Quasinormal modes in terms of Null Geodesic Parameters}\label{qnm_from_null}

In order to find an approximate analytic expression for the quasinormal modes in the asymptotic limit $l\rightarrow \infty$, 
we drop the higher order terms in (\ref{6th_order_formula}) and write

\begin{equation}\label{6th_order_formula_asymp}
 \frac{Q_0}{\sqrt{2Q_0^{''}}}=i\left(n+\frac{1}{2}\right).
\end{equation}

It can be seen that in the limit $l\rightarrow \infty$, the effective potentials $V(r)$ for all three types of perturbations, 
given by (\ref{effective_potentials}), reduce to much simpler forms so that simple expressions are obtained for the corresponding 
functions $Q_0$ as follows:

\begin{equation}\label{Q_0_asymp}
 Q_0 \simeq \Omega^2-Cl^2\frac{f}{r^2},
\end{equation}

where the values of the parameter $C$ for tensor ($C_t$), vector ($C_v$) and scalar ($C_s$) perturbations in $d$ dimensions for 
the Lovelock theory of order $k$ take the form:

\begin{equation}
 C=\left\{
\begin{array}{ll}
C_{t}=\frac{1}{d-4}\Biggl[(d-4)-(k-1)\left(\frac{d-1}{k}\right)\Biggr] \\ 
\\
C_v=\frac{1}{d-3}\Biggl[(d-3)-(k-1)\left(\frac{d-1}{k}\right)\Biggr] \\ 
\\
C_s=\frac{1}{d-2}\Biggl[(d-2)-(k-1)\left(\frac{d-1}{k}\right)\Biggr] \label{constants}
\end{array}
\right.
\end{equation}

Substituting (\ref{constants}) and (\ref{Q_0_asymp}) into (\ref{6th_order_formula_asymp}), we get the following expression for the 
quasinormal modes in the limit $l\rightarrow \infty$:

\begin{equation}\label{Omega_asymp}
 \Omega_{asy}=l\sqrt{C}\sqrt{\frac{f_c}{r_c^2}}
-i\frac{\left(n+\frac{1}{2}\right)}{\sqrt{2}}
\sqrt{-\frac{r_c^2}{f_c}\left[\frac{d^2}{dr_*^2}\left(\frac{f}{r^2}\right)\right]_{r=r_c}},
\end{equation}\\

with $C$ taking appropriate values depending on the type of perturbation under consideration. The connection between $\Omega_{asy}$ 
and the null geodesic parameters is clear from (\ref{omega_c}), (\ref{lyapunov}) and (\ref{Omega_asymp}). Clearly, the real 
parts of the modes vary linearly with $l$ while the imaginary parts are independent of $l$. Thus, for the same value of $n$, the imaginary 
parts of the modes should approach a constant. Also, given sufficiently high value of the parameter $d$, we have 
$C_t\simeq C_v\simeq C_s$, which means that the metric perturbations of the spacetime given by (\ref{metric}) should be isospectral 
if one considers Lovelock theories given by (\ref{llaction}) in very high dimensions.  

\section{Results and Discussion}\label{discussion}

We use (\ref{6th_order_formula}) to compute the QNMs 
$\Omega$ for various combinations 
of spacetime dimension $d$ and the order parameter $k$. 
The calculation is done for different values of the mode number $n$. We have tabulated the low-lying modes for $l=2$ in Tables \ref{table_tensor_fundamental}, 
\ref{table_vector_fundamental} and \ref{table_scalar_fundamental}. The parameter $l$ is given values from 
6 to 80 and selected values of the QNMs are tabulated in tables \ref{table_tensor}, \ref{table_vector} and 
\ref{table_scalar}. Tables \ref{table_tensor_vs_k}, \ref{table_vector_vs_k} and \ref{table_scalar_vs_k} 
show the QNMs for various values of the order $k$. In Table \ref{table_comparison}, we compare the 
values of QNMs obtained using the eikonal approximation and the sixth order WKB method. 
In all tables and figures in this work, $\omega$ stands for $\Omega G_kM$, 
where $\Omega$ is the QNM calculated using (\ref{6th_order_formula}).\\

\begin{figure}[h]
\centering
\includegraphics[width=0.65\columnwidth]{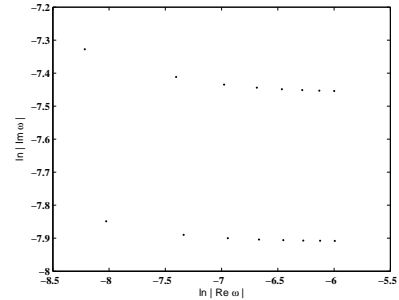}
\caption{Tensor modes for $k=2$ and $d=8$, for $n=5$ (top) and $n=3$ (bottom). The plotted points within each 
curve are for $l=10$ (left) to $l=80$ (right). } \label{scatter_tensor_modes}
\end{figure} 

\begin{figure}[h]
\centering
\includegraphics[width=0.65\columnwidth]{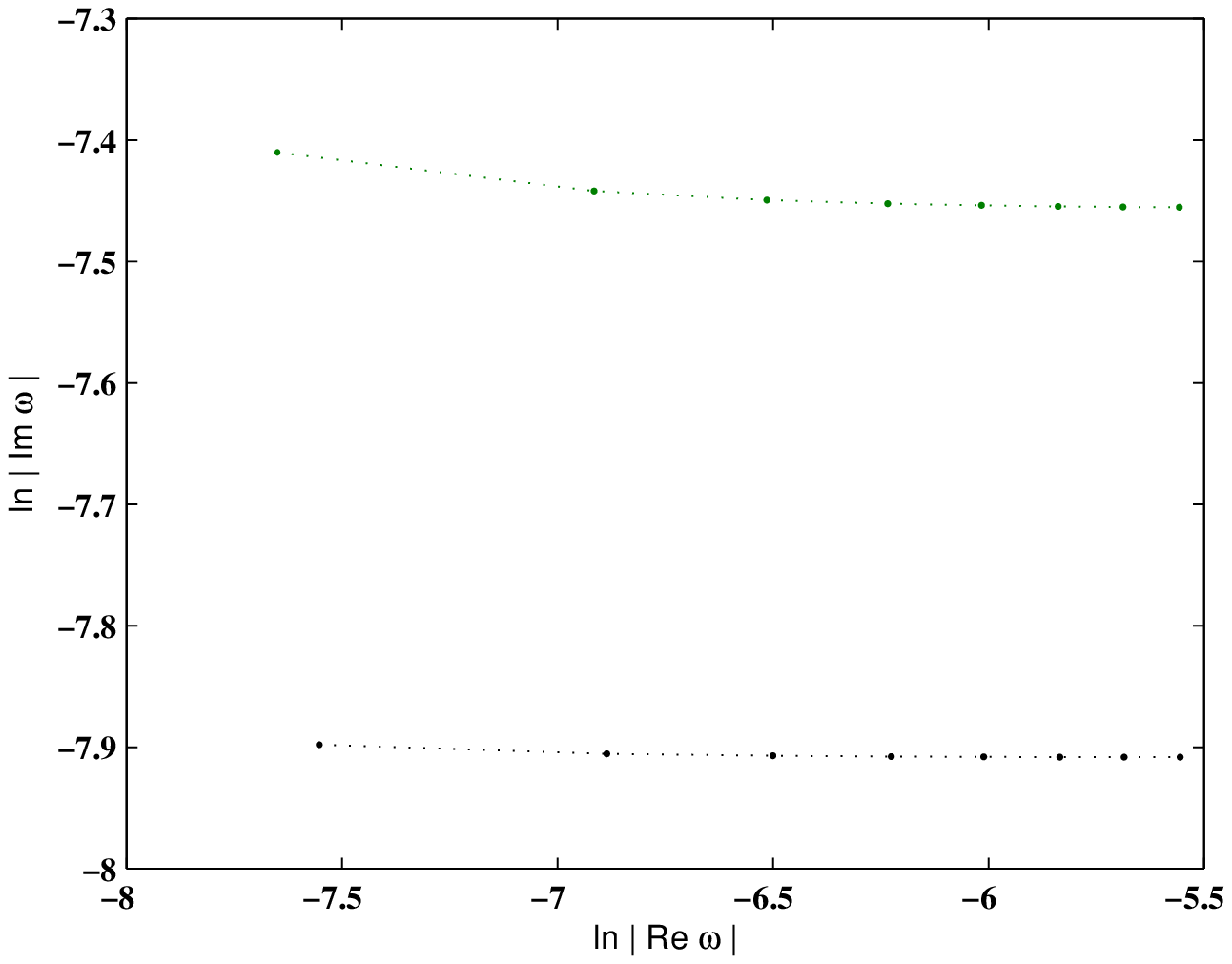}
\caption{Vector modes for $k=2$ and $d=8$, for $n=5$ (top) and $n=3$ (bottom). The plotted points within each 
curve are for $l=10$ (left) to $l=80$ (right). } \label{scatter_vector_modes}
\end{figure} 

\begin{figure}[h]
\centering
\includegraphics[width=0.65\columnwidth]{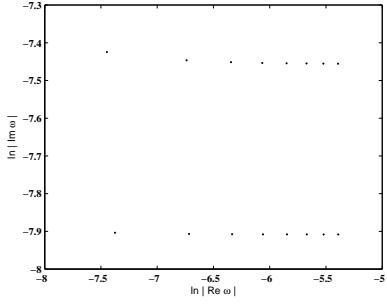}
\caption{Scalar modes for $k=2$ and $d=8$, for $n=5$ (top) and $n=3$ (bottom). The plotted points within each 
curve are for $l=10$ (left) to $l=80$ (right). } \label{scatter_scalar_modes}
\end{figure} 
  
Figs. \ref{scatter_tensor_modes}, \ref{scatter_vector_modes} and \ref{scatter_scalar_modes} are $log-log$ plots of 
the QNMs for tensor, vector and scalar modes respectively, which show the behavior of the modes as the parameter $l$ varies 
from relatively low values to high values. 
From the plots, we observe a behavior that is consistent with that suggested by the 
null geodesic method.  We see that 
the the imaginary parts of the modes tend to become a costant at high values of $l$, just as suggested by (\ref{Omega_asymp}). 
The behaviour of the imaginary parts for lower values of $l$ is similar to that in an earlier work \cite{chen_wang} 
which also shows a convergent pattern for $Im~\omega$ as $l$ increases.
\\

\begin{figure}[h]
\centering
\includegraphics[width=0.65\columnwidth]{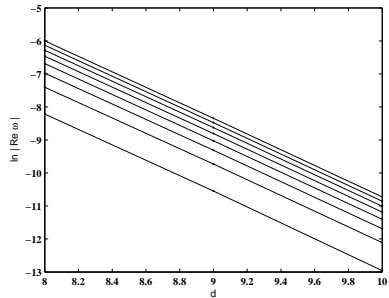}
\caption{Variation of $ln~|Re(\omega)|$ vs $d$ for $k=2$ for Tensor modes . Here, $n=5$. The curves are for $l=10$ (bottom) 
to $l=50$ (top). } \label{tensor_re_w_vs_d}
\end{figure}

\begin{figure}[h]
\centering
\includegraphics[width=0.65\columnwidth]{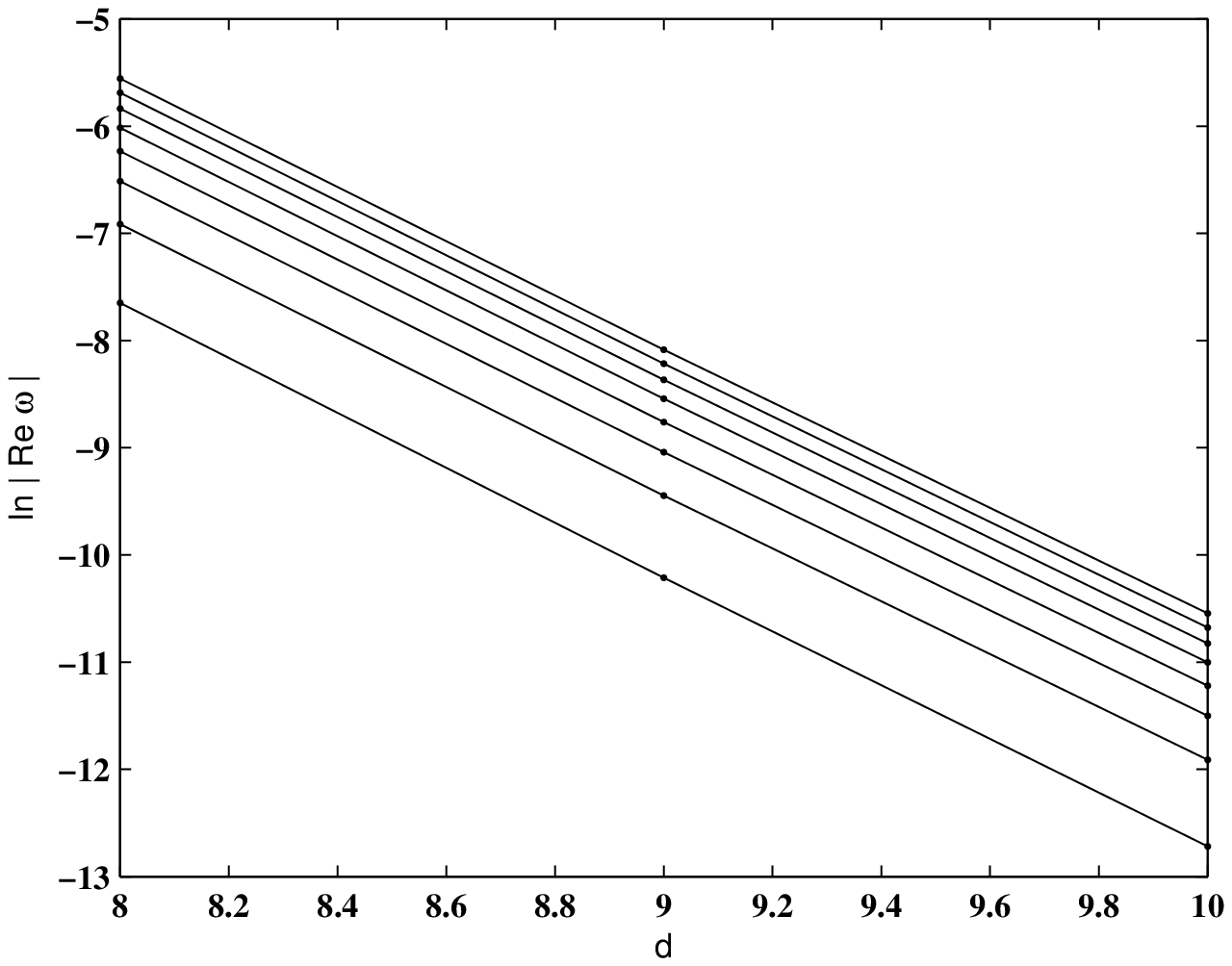}
\caption{Variation of $ln~|Re(\omega)|$ vs $d$ for $k=2$ for Vector modes . Here, $n=5$. The curves are for $l=10$ (bottom) 
to $l=50$ (top). } \label{vector_re_w_vs_d}
\end{figure}

\begin{figure}[h]
\centering
\includegraphics[width=0.65\columnwidth]{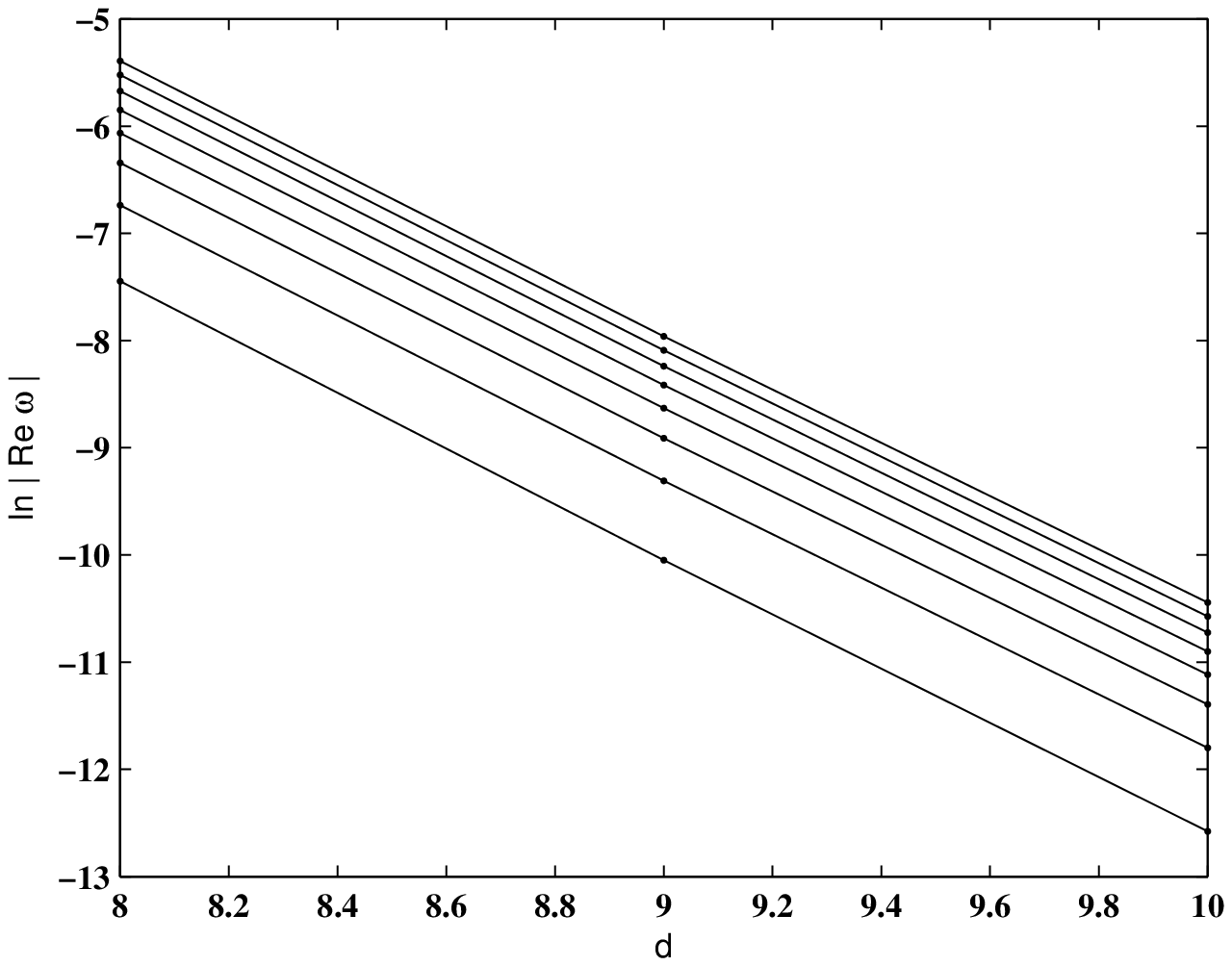}
\caption{Variation of $ln~|Re(\omega)|$ vs $d$ for $k=2$ for Scalar modes . Here, $n=5$. The curves are for $l=10$ (bottom) 
to $l=50$ (top). } \label{scalar_re_w_vs_d}
\end{figure}

Figs. \ref{tensor_re_w_vs_d} to \ref{scalar_re_w_vs_d} show the variation of 
logarithm of the the absolute values of the real parts of the QNMs with spacetime dimension $d$. As 
observed from the plots, the real parts decrease as $d$ increases, indicating modes with lower frequency 
in higher dimensions. For any value of $d$, the real parts 
increase with increasing values of $l$.\\

\begin{figure}[h]
\centering
\includegraphics[width=0.65\columnwidth]{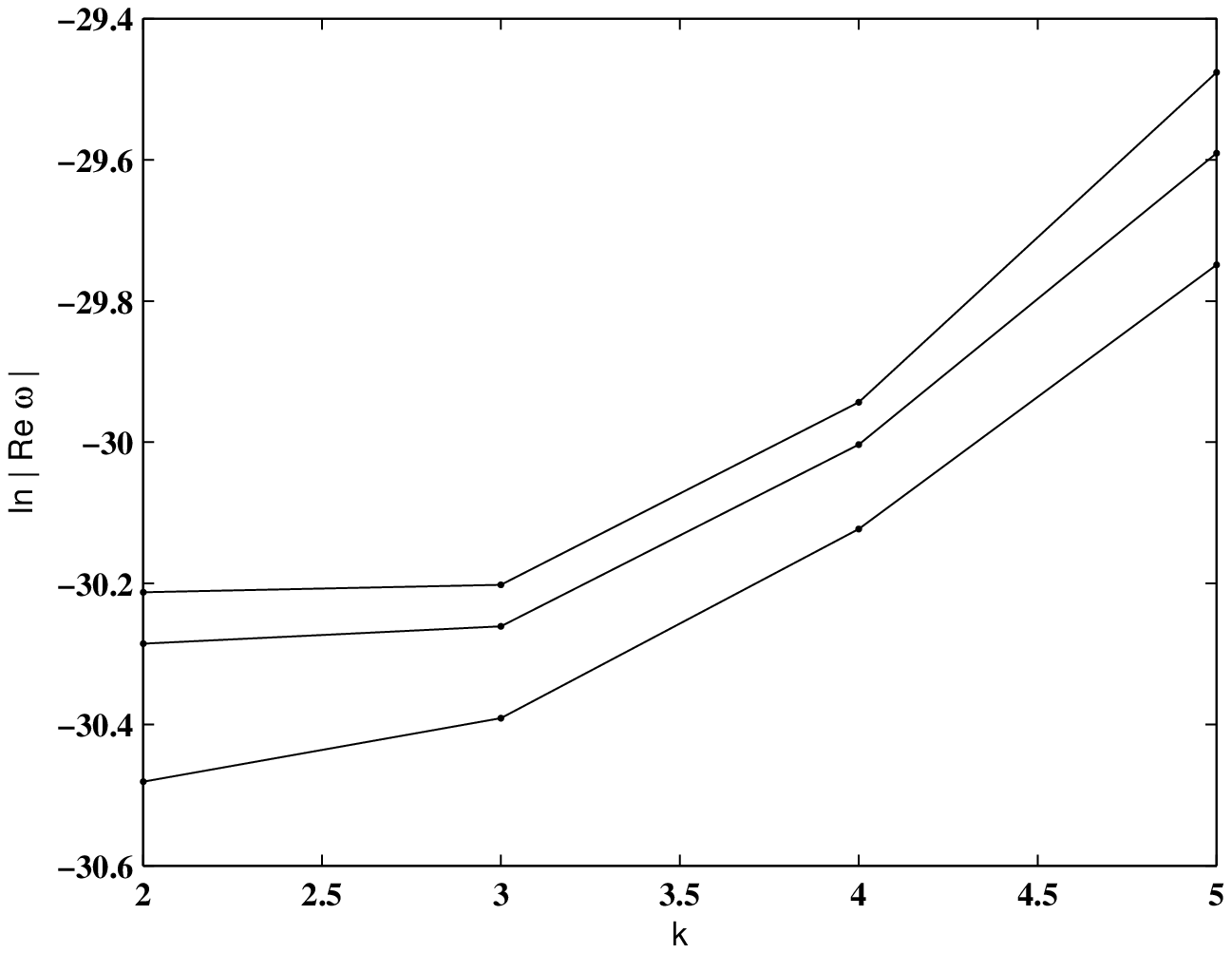}
\caption{Variation of $ln~|Re(\omega)|$ vs $k$ for $d=17$ and $l=7$ for Tensor modes . The curves are for $n=0$ (top) 
to $n=2$ (bottom). } \label{tensor_Re_w_vs_k}
\end{figure}

\begin{figure}[h]
\centering
\includegraphics[width=0.65\columnwidth]{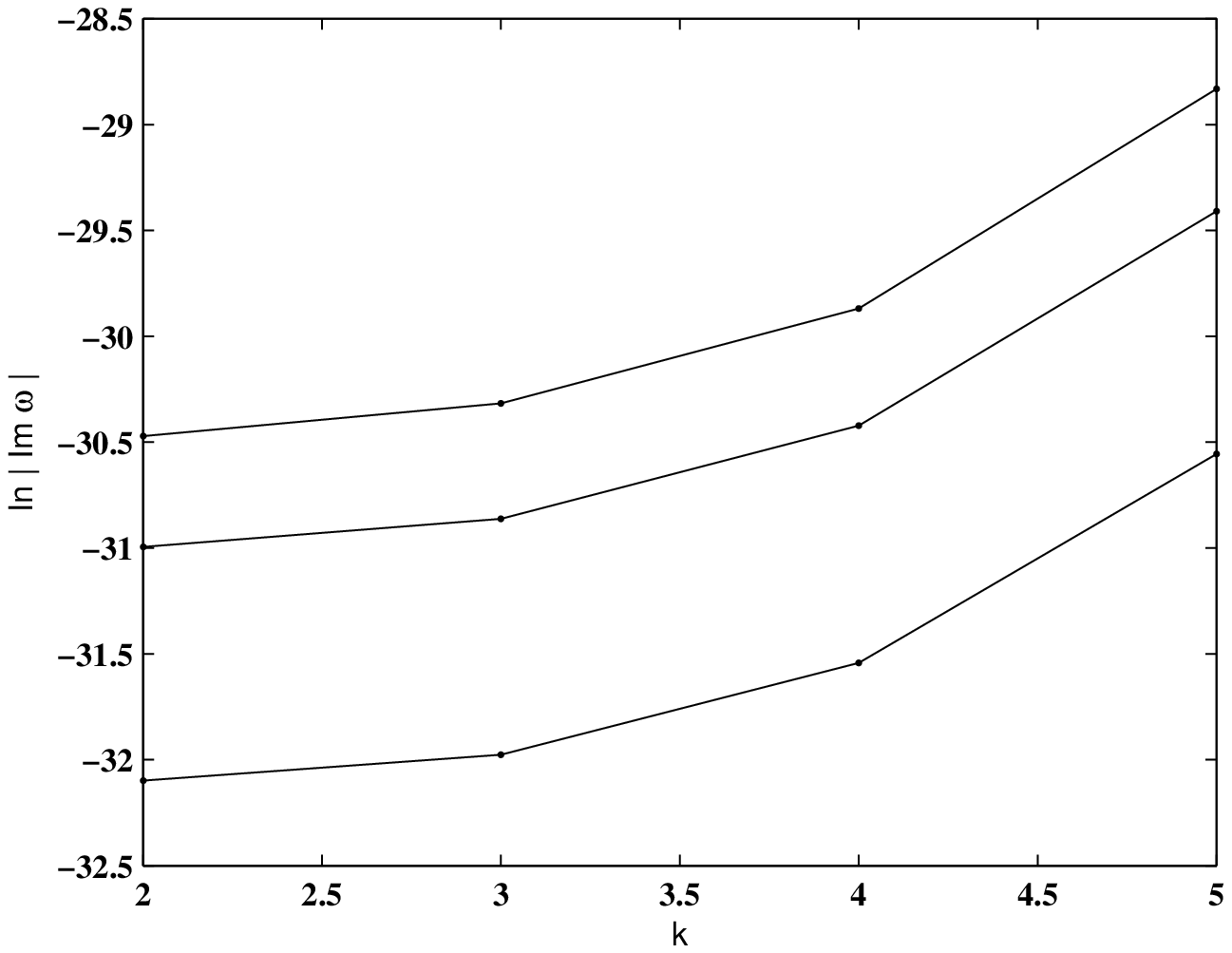}
\caption{Variation of $ln~|Im(\omega)|$ vs $k$ for $d=17$ and $l=7$ for Tensor modes . The curves are for $n=0$ (bottom) 
to $n=2$ (top). } \label{tensor_Im_w_vs_k}
\end{figure}

\begin{figure}[h]
\centering
\includegraphics[width=0.65\columnwidth]{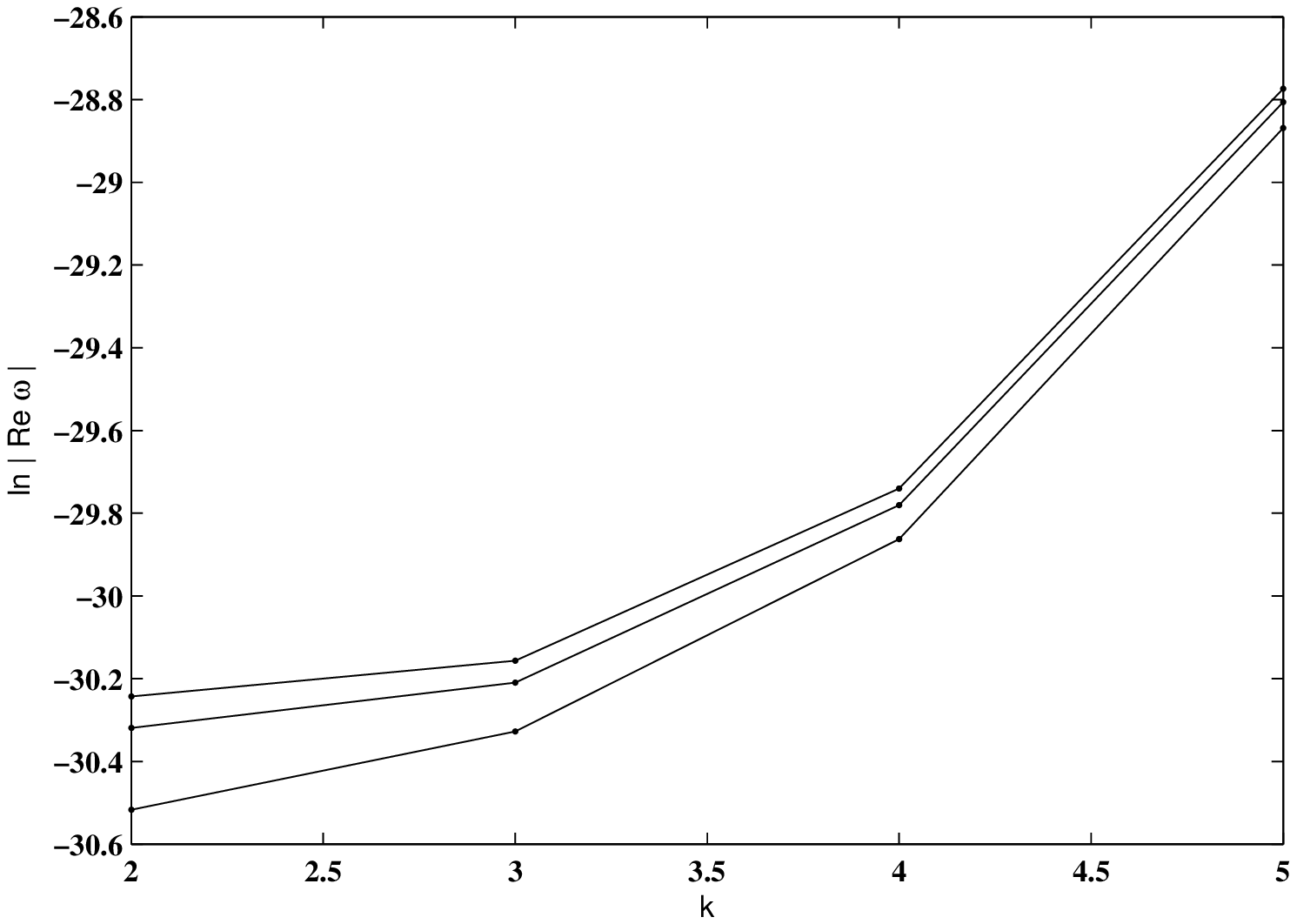}
\caption{Variation of $ln~|Re(\omega)|$ vs $k$ for $d=17$ and $l=7$ for Vector modes . The curves are for $n=0$ (top) 
to $n=2$ (bottom). } \label{vector_Re_w_vs_k}
\end{figure}

\begin{figure}[h]
\centering
\includegraphics[width=0.65\columnwidth]{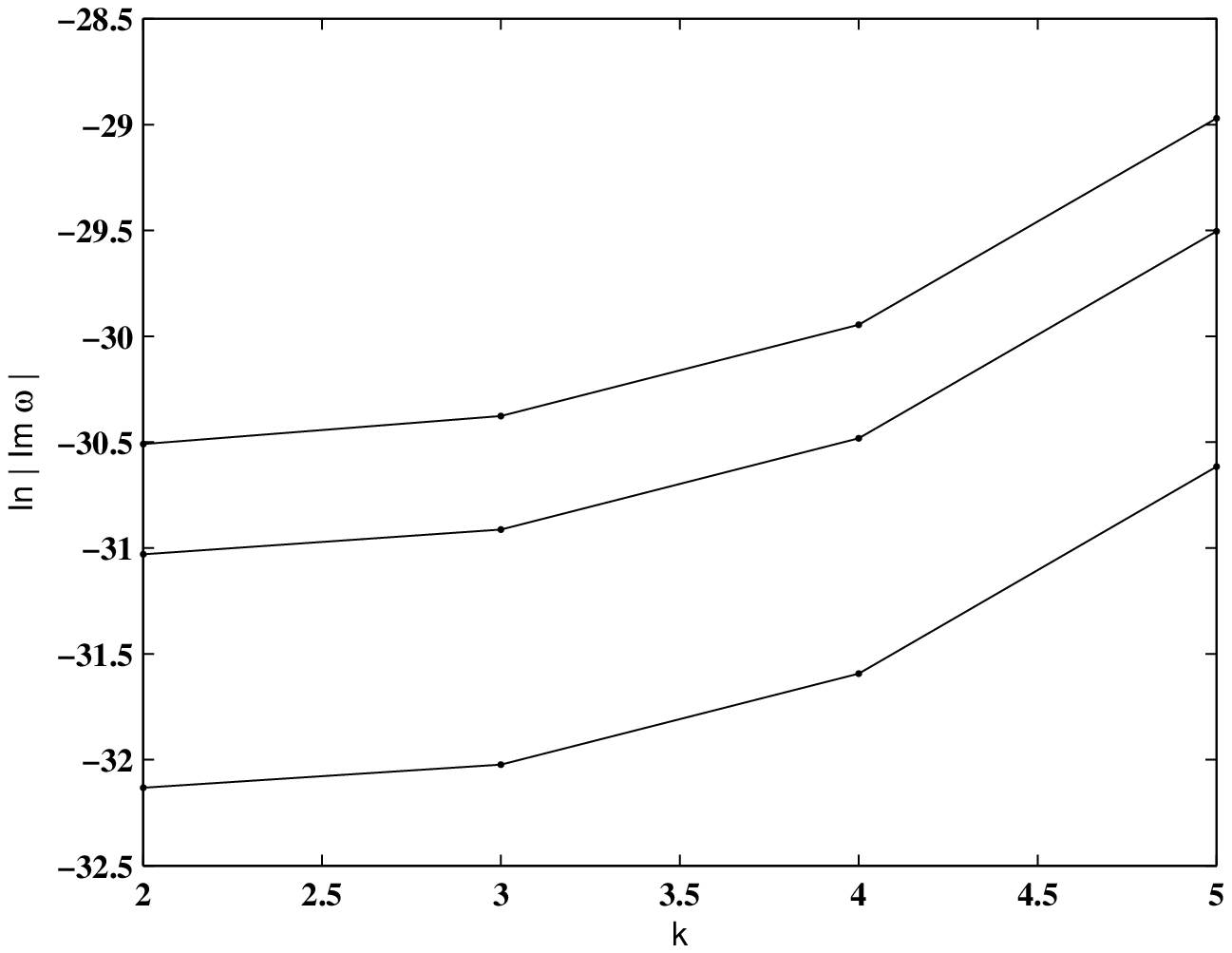}
\caption{Variation of $ln~|Im(\omega)|$ vs $k$ for $d=17$ and $l=7$ for Vector modes . The curves are for $n=0$ (bottom) 
to $n=2$ (top). } \label{vector_Im_w_vs_k}
\end{figure}

\begin{figure}[h]
\centering
\includegraphics[width=0.65\columnwidth]{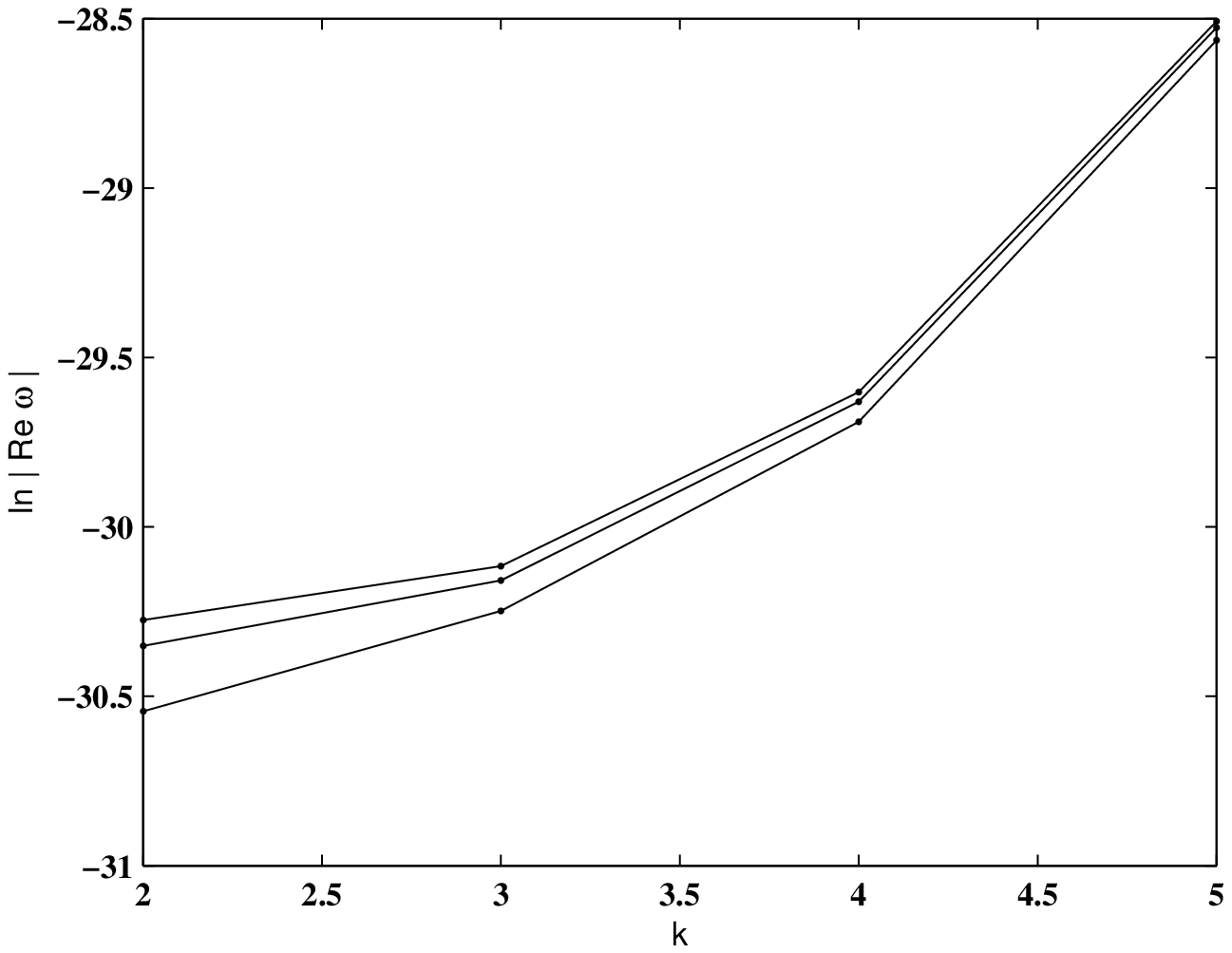}
\caption{Variation of $ln~|Re(\omega)|$ vs $k$ for $d=17$ and $l=7$ for Scalar modes . The curves are for $n=0$ (top) 
to $n=2$ (bottom). } \label{scalar_Re_w_vs_k}
\end{figure}

\begin{figure}[h]
\centering
\includegraphics[width=0.65\columnwidth]{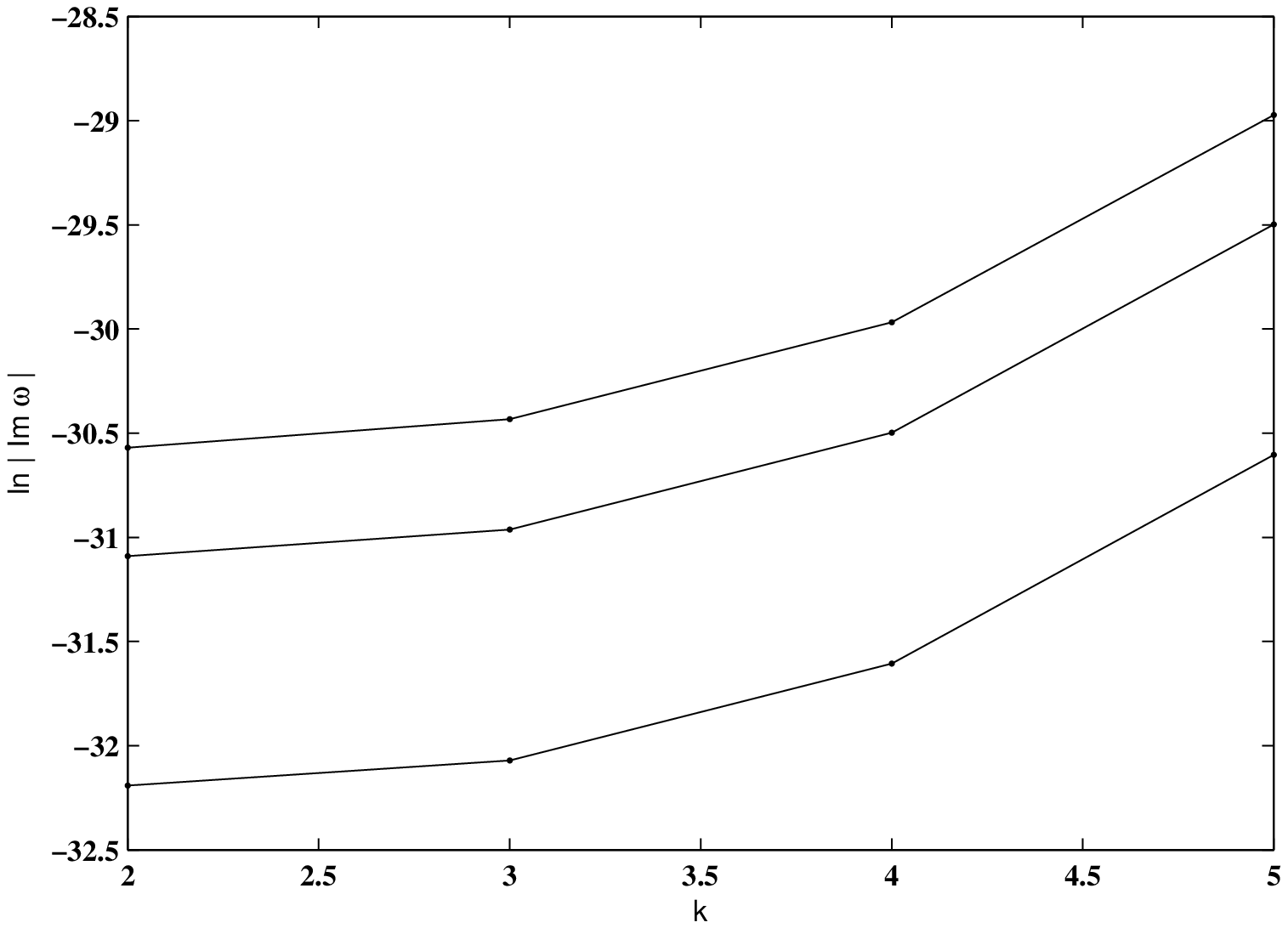}
\caption{Variation of $ln~|Im(\omega)|$ vs $k$ for $d=17$ and $l=7$ for Scalar modes . The curves are for $n=0$ (bottom) 
to $n=2$ (top). } \label{scalar_Im_w_vs_k}
\end{figure}

Figs. \ref{tensor_Re_w_vs_k} to \ref{scalar_Im_w_vs_k} show the variation of 
logarithm of the the absolute values of the real and imaginary parts of the QNMs with the order parameter $k$. As 
observed from the plots, the real parts as well as the imaginary parts increase as $k$ increases.\\

\begin{table*}[ph]
\caption{\label{table_tensor_fundamental}Low-lying modes for Tensor perturbations for various dimensions (in units of $10^{-6}$)}

\begin{tabular}{cccccc}
\\ \hline 
\hline \\\\
$l$  &  $n$  &  $\omega\ (d=8,k=2)$  &  $\omega\ (d=9,k=2)$  &  $\omega\ (d=10,k=2)$ \\\\ \hline \hline
\\

2 & 0 & 133.1406 - 54.2445$i$ & 13.4214 - 4.7089$i$ & 1.3217 - 0.4333$i$\\

2 & 1 & 112.3407 - 173.8648$i$ & 11.3954 - 15.0050$i$ & 1.1125 - 1.3804$i$\\

2 & 2 & 91.9167 - 316.0219$i$ & 8.5470 - 27.6200$i$ & 0.7721 - 2.5675$i$\\\\

3 & 0 & 164.0819 - 53.5968$i$ & 16.4993 - 4.6478$i$ & 1.6117 - 0.4271$i$\\

3 & 1 & 145.2940 - 168.7927$i$ & 14.7436 - 14.5361$i$ & 1.4336 - 1.3333$i$\\

3 & 2 & 120.0297 - 303.0023$i$ & 11.8915 - 26.0985$i$ & 1.1173 - 2.4025$i$\\

3 & 3 & 103.2647 - 458.7079$i$ & 8.8570 - 40.1800$i$ & 0.7309 - 3.7512$i$\\\\

4 & 0 & 194.8473 - 53.2421$i$ & 19.5138 - 4.6210$i$ & 1.8935 - 0.4247$i$\\

4 & 1 & 178.1521 - 165.5511$i$ & 17.9892 - 14.2800$i$ & 1.7400 - 1.3099$i$\\

4 & 2 & 152.4328 - 292.9591$i$ & 15.3267 - 25.1695$i$ & 1.4547 - 2.3105$i$\\

4 & 3 & 128.2170 - 439.3342$i$ & 12.1695 - 37.9506$i$ & 1.0840 - 3.5106$i$\\

4 & 4 & 114.7935 - 606.6123$i$ & 9.1155 - 53.2629$i$ & 0.6793 - 4.9998$i$\\

\\
\hline
 
\end{tabular}
\end{table*}

\begin{table*}[ph]
\caption{\label{table_vector_fundamental}Low-lying modes for Vector perturbations for various dimensions (in units of $10^{-6}$)}

\begin{tabular}{cccccc}
\\ \hline 
\hline \\\\
$l$  &  $n$  &   $\omega\ (d=7,k=2)$ &  $\omega\ (d=8,k=2)$ &  $\omega\ (d=9,k=2)$  &  $\omega\ (d=10,k=2)$ \\\\ \hline \hline
\\

2 & 0 & 1987.1950 - 730.1241$i$ & 134.2193 - 50.3052$i$ & 11.9857 - 4.4229$i$ & -\\

2 & 1 & 1608.5488 - 2369.9620$i$ & 108.6194 - 161.0435$i$ & 9.5496 - 14.0856$i$ & -\\

2 & 2 & 1060.5751 - 4440.9963$i$ & 66.9691 - 302.4766$i$ & 5.2406 - 26.7137$i$ & -\\\\

3 & 0 & 3062.5553 - 728.6387$i$ & 196.1230 - 48.6493$i$ & 16.7160 - 4.2920$i$ & 1.5269 - 0.4004$i$\\

3 & 1 & 2838.5908 - 2241.7117$i$ & 176.6061 - 150.7660$i$ & 14.6955 - 13.4016$i$ & 1.3218 - 1.2518$i$\\

3 & 2 & 2446.9666 - 3914.5570$i$ & 141.2707 - 267.9013$i$ & 11.0103 - 24.1834$i$ & 0.9361 - 2.2729$i$\\

3 & 3 & 1981.3414 - 5816.5456$i$ & 97.0424 - 409.3032$i$ & 6.3318 - 37.7686$i$ & 0.4302 - 3.5975$i$\\\\

4 & 0 & 4014.3270 - 748.3938$i$ & 253.8493 - 49.5175$i$ & 21.3443 - 4.3038$i$ & 1.9199 - 0.3974$i$\\

4 & 1 & 3846.5184 - 2282.0116$i$ & 239.8692 - 151.2782$i$ & 19.8584 - 13.1766$i$ & 1.7586 - 1.2201$i$\\

4 & 2 & 3544.7474 - 3920.3525$i$ & 213.8152 - 261.3232$i$ & 16.9958 - 22.9209$i$ & 1.4407 - 2.1399$i$\\

4 & 3 & 3169.0262 - 5709.3996$i$ & 179.5279 - 384.6027$i$ & 13.0678 - 34.2095$i$ & 0.9950 - 3.2436$i$\\

4 & 4 & 2783.4865 - 7671.34073$i$ & 141.2937 - 525.2555$i$ & 8.4835 - 47.7143$i$ & 0.4665 - 4.6197$i$\\

\\
\hline
 
\end{tabular}
\end{table*}

\begin{table*}[ph]
\caption{\label{table_scalar_fundamental}Low-lying modes for Scalar perturbations for various dimensions (in units of $10^{-6}$)}

\begin{tabular}{cccccc}
\\ \hline 
\hline \\\\
$l$  &  $n$  &   $\omega\ (d=7,k=2)$  &  $\omega\ (d=8,k=2)$  &  $\omega\ (d=9,k=2)$  &  $\omega\ (d=10,k=2)$ \\\\ \hline \hline
\\

2 & 0 & 2547.5661 - 716.4415$i$ & 205.36410 - 9.6575$i$ & - & 2.3027 - 0.3208$i$\\

2 & 1 & 2271.1626 - 2240.5828$i$ & 334.0324 - 6.4720$i$ & - & 6.9142 - 0.4909$i$\\

2 & 2 & 1810.0458 - 3995.1437$i$ & 90.0530 - 763.3169$i$ & - & 15.6242 - 0.5226$i$\\\\

3 & 0 & 3905.5647 - 746.6365$i$ & 218.9954 - 46.7773$i$ & 16.9439 - 3.9805$i$ & 1.5315 - 0.2688$i$\\

3 & 1 & 3733.1930 - 2278.6463$i$ & 204.5237 - 142.9336$i$ & 14.9968 - 13.7331$i$ & 1.9261 - 0.4939$i$\\

3 & 2 & 3424.6745 - 3920.6131$i$ & 177.8404 - 246.4931$i$ & 12.4106 - 29.1985$i$ & 3.5782 - 0.1568$i$\\

3 & 3 & 3043.6602 - 5720.0577$i$ & 142.4973 - 361.3841$i$ & 10.8457 - 52.7568$i$ & -\\\\

4 & 0 & 5111.2845 - 761.7854$i$ & 290.4033 - 48.9992$i$ & 22.7621 - 4.1094$i$ & 1.9517 - 0.3631$i$\\

4 & 1 & 4978.7417 - 2309.6889$i$ & 278.8390 - 149.1029$i$ & 21.5711 - 12.5907$i$ & 1.8526 - 1.1095$i$\\

4 & 2 & 4732.6146 - 3927.8632$i$ & 257.1089 - 255.5290$i$ & 19.3001 - 21.9070$i$ & 1.6680 - 1.9276$i$\\

4 & 3 & 4408.8671 - 5651.5157$i$ & 228.0448 - 372.0127$i$ & 16.2394 - 32.6401$i$ & 1.4339 - 2.8909$i$\\

4 & 4 & 4051.0515 - 7500.0861$i$ & 195.0275 - 501.4720$i$ & 12.7435 - 45.3070$i$ & 1.2037 - 4.0923$i$\\

\\
\hline
 
\end{tabular}
\end{table*}

\begin{table*}[ph]
\caption{\label{table_tensor}QNMs of Tensor perturbations for $n=1$ (in units of $10^{-6}$)}

\begin{tabular}{cccc}
\hline 
\hline \\\\

 $l$  &  $\omega\ (d=8,k=2)$  &  $\omega\ (d=9,k=2)$  &  $\omega\ (d=10,k=2)$ \\\\ \hline \\

10 & 368.5710 - 159.6868$i$ & 36.3453 - 13.8559$i$ & 3.4472 - 1.2732$i$\\

20 & 676.4250 - 158.2275$i$ & 65.7576 - 13.7504$i$ & 6.1611 - 1.2638$i$\\

30 & 981.4976 - 157.8903$i$ & 94.8519 - 13.7256$i$ & 8.8397 - 1.2616$i$\\

40 & 1285.7250 - 157.7616$i$ & 123.8498 - 13.7161$i$ & 11.5075 - 1.2607$i$\\

50 & 1589.5868 - 157.6991$i$ & 152.8060 - 13.7114$i$ & 14.1705 - 1.2603$i$\\

60 & 1893.2578 - 157.6642$i$ & 181.7402 - 13.7088$i$ & 16.8310 - 1.2600$i$\\

70 & 2196.8167 - 157.6427$i$ & 210.6616 - 13.7071$i$ & 19.4901 - 1.2599$i$\\\\
\hline
 
\end{tabular}
\end{table*}

\begin{table*}[ph]
\caption{\label{table_vector}QNMs of Vector perturbations for $n=1$ (in units of $10^{-6}$)}
\begin{tabular}{cccc}
\hline 
\hline \\\\
$l$  &  $\omega\ (d=8,k=2)$  &  $\omega\ (d=9,k=2)$  &  $\omega\ (d=10,k=2)$ \\\\ \hline \\

10 & 554.9965 - 155.8556$i$ & 45.3264 - 13.5203$i$ & 3.9634 - 1.2402$i$\\

20 & 1039.3543 - 157.05278$i$ & 84.0122 - 13.6437$i$ & 7.2757 - 1.2529$i$\\

30 & 1514.6547 - 157.0527$i$ & 121.8428 - 13.6738$i$ & 10.5038 - 1.2562$i$\\

40 & 1987.3587 - 157.4331$i$ & 159.4219 - 13.6856$i$ & 13.7066 - 1.2575$i$\\

50 & 2458.9581 - 157.4840$i$ & 196.8927 - 13.6913$i$ & 16.8984 - 1.2582$i$\\

60 & 2929.9849 - 157.5124$i$ & 234.3070 - 13.6946$i$ & 20.0845 - 1.2585$i$\\

70 & 3400.6768 - 157.5299$i$ & 271.6881 - 13.6966$i$ & 23.2670 - 1.2588$i$\\\\
\hline
 
\end{tabular}
\end{table*}

\begin{table*}[ph]
\caption{\label{table_scalar}QNMs of Scalar perturbations for $n=1$ (in units of $10^{-6}$)}
\begin{tabular}{cccc}
\hline 
\hline \\\\
$l$  &  $\omega\ (d=8,k=2)$  &  $\omega\ (d=9,k=2)$  &  $\omega\ (d=10,k=2)$ \\\\ \hline \\

10 & 652.8312 - 155.8358$i$ & 50.9723 - 13.4597$i$ & 4.3254 - 1.2285$i$\\

20 & 1224.3284 - 157.0746$i$ & 95.0437 - 13.6313$i$ & 8.0203 - 1.2502$i$\\

30 & 1784.6681 - 157.3415$i$ & 138.0091 - 13.6685$i$ & 11.6034 - 1.2549$i$\\

40 & 2341.8477 - 157.4900$i$ & 180.6552 - 13.6826$i$ & 15.1536 - 1.2568$i$\\

50 & 2897.6879 - 157.4900$i$ & 223.1647 - 13.6894$i$ & 18.6895 - 1.2577$i$\\

60 & 3452.8349 - 157.5165$i$ & 265.6032 - 13.6932$i$ & 22.2179 - 1.2582$i$\\

70 & 4007.5767 - 157.5330$i$ & 307.9999 - 13.6956$i$ & 25.7420 - 1.2585$i$\\\\
\hline
 
\end{tabular}
\end{table*}

\begin{table*}[ph]
\caption{\label{table_comparison}Comparison between the eikonal approx. and the numerical values of QNMs ($d=10$, $k=2$ and $n=1$ ) (in units of $10^{-6}$)}

\begin{tabular}{ccccccc}
\hline 
\hline \\\\

\\&\multicolumn{2}{c}{Tensor}&\multicolumn{2}{c}{Vector}&\multicolumn{2}{c}{Scalar}\\\\\hline\\

 $l$  &  $\omega_{eik}$  &  $\omega_{num}$  &  $\omega_{eik}$  &  $\omega_{num}$ &  $\omega_{eik}$  &  $\omega_{num}$ \\\\ \hline \\

10 & 2.65511 - 1.34949$i$ & 3.4472 - 1.2732$i$ & 3.17346 - 1.34949$i$ & 3.9634 - 1.2402$i$ & 3.51238 - 1.34949$i$ & 4.3254 - 1.2285$i$\\

20 & 5.31022 - 1.34949$i$ & 6.1611 - 1.2638$i$ & 6.34692 - 1.34949$i$ & 7.2757 - 1.2529$i$ & 7.02476 - 1.34949$i$ & 8.0203 - 1.2502$i$\\

30 & 7.96532 - 1.34949$i$ & 8.8397 - 1.2616$i$ & 9.52038 - 1.34949$i$ & 10.5038 - 1.2562$i$ & 10.5371 - 1.34949$i$ & 11.6034 - 1.2549$i$\\

40 & 10.6204 - 1.34949$i$ & 11.5075 - 1.2607$i$ & 12.6938 - 1.34949$i$ & 13.7066 - 1.2575$i$ & 14.0495 - 1.34949$i$ & 15.1536 - 1.2568$i$\\

50 & 13.2755 - 1.34949$i$ & 14.1705 - 1.2603$i$ & 15.8673 - 1.34949$i$ & 16.8984 - 1.2582$i$ & 17.5619 - 1.34949$i$ & 18.6895 - 1.2577$i$\\

\hline
 
\end{tabular}
\end{table*}

\begin{table*}[ph]
\caption{\label{table_tensor_vs_k}QNMs for Tensor perturbations for various values of $k$ with $d=17$ and $l=7$ (in units of $10^{-14}$)}

\begin{tabular}{ccccc}
\\ \hline 
\hline \\\\
$n$  &  $\omega\ (k=2)$  &  $\omega\ (k=3)$  &  $\omega\ (k=4)$ &$\omega\ (k=5)$\\\\ \hline \hline
\\

0 & 7.5654 - 1.1475$i$ & 7.6440 - 1.2976$i$ & 9.9011 - 2.0021$i$ & 15.8018 - 5.3654$i$\\

1 & 7.0333 - 3.4599$i$ & 7.2096 - 3.9518$i$ & 9.3245 - 6.1374$i$ & 14.0911 - 16.8858$i$\\

2 & 5.7838 - 5.8365$i$ & 6.3300 - 6.8135$i$ & 8.2728 - 10.6688$i$ & 12.0281 - 30.0839$i$\\\\

\\
\hline
 
\end{tabular}
\end{table*}

\clearpage
\begin{table*}[ph]
\caption{\label{table_vector_vs_k}QNMs for Vector perturbations for various values of $k$ with $d=17$ and $l=7$ (in units of $10^{-14}$)}

\begin{tabular}{ccccc}
\\ \hline 
\hline \\\\
$n$  &  $\omega\ (k=2)$  &  $\omega\ (k=3)$  &  $\omega\ (k=4)$ &$\omega\ (k=5)$\\\\ \hline \hline
\\

0 & 7.3405 - 1.1101$i$ & 8.0021 - 1.2382$i$ & 12.1344 - 1.9030$i$ & 31.8981 - 5.0575$i$\\

1 & 6.8017 - 3.3424$i$ & 7.5894 - 3.7559$i$ & 11.6583 - 5.7797$i$ & 30.8876 - 15.3589$i$\\

2 & 5.5808 - 5.6277$i$ & 6.7427 - 6.4196$i$ & 10.7347 - 9.8779$i$ & 29.0064 - 26.2060$i$\\\\

\\
\hline
 
\end{tabular}
\end{table*}

\begin{table*}[ph]
\caption{\label{table_scalar_vs_k}QNMs for Scalar perturbations for various values of $k$ with $d=17$ and $l=7$ (in units of $10^{-14}$)}

\begin{tabular}{ccccc}
\\ \hline 
\hline \\\\
$n$  &  $\omega\ (k=2)$  &  $\omega\ (k=3)$  &  $\omega\ (k=4)$ &$\omega\ (k=5)$\\\\ \hline \hline
\\

0 & 7.1050 - 1.0462$i$ & 8.3319 - 1.1800$i$ & 13.9275 - 1.8777$i$ & 41.6200 - 5.1162$i$\\

1 & 6.5859 - 3.1448$i$ & 7.9915 - 3.5720$i$ & 13.5324 - 5.6876$i$ & 40.8428 - 15.4635$i$\\

2 & 5.4291 - 5.2934$i$ & 7.3036 - 6.0721$i$ & 12.7618 - 9.6663$i$ & 39.3575 - 26.1500$i$\\\\

\\
\hline
 
\end{tabular}
\end{table*}

\section{Conclusion}\label{conclusion}

In summary, we have studied the quasinormal modes of metric perturbations of tensor, vector and scalar type for 
asymptotically flat black hole spacetimes for a particular class of theories in the Lovelock 
model. These theories are specified by the action given by (\ref{llaction}) with the higher order coupling constants 
given by (\ref{alphas}). We used the sixth order WKB formula for the quasinormal modes \cite{konoplya_order6}
 in order to compute the QNMs for various values of $d$ and $k$. We also used the 
connection between null geodesic parameters and the asymptotic quasinormal modes of static and spherically 
symmetric spacetimes, established in \cite{cardoso_miranda_berti}, to deduce an analytic form for the asymptotic 
modes in the limit $l \rightarrow \infty$. Numerical analysis indicates that 
the asymptotic behavior of the QNMs in higher ordered theories is indeed consistent with the theory, as can be seen easily from Table \ref{table_comparison}. 
We observe that the imaginary parts of the modes attain a constant value for very high values of the parameter $l$, just 
as suggested by the null geodesic method. We calculated the quasinormal modes of perturbations for different orders of the Lovelock theory 
and found that the real as well as imaginary parts of the modes increase with increasing values of $k$. 
We also find that the real parts of the modes decrease with increase in the 
spacetime dimension $d$. The theory also suggests that the modes should be approximately isospectral 
at high values of $d$. This is seen to hold roughly at $d\geq10$, especially in the case of imaginary parts.
 The quasinormal behavior revealed in this study helps us understand better the dynamics of fields in the 
vicinity of black holes in higher ordered theories of gravity.

\section{Acknowledgment}\label{ack}

CBP would like to acknowledge financial assistance from UGC, New Delhi through the UGC-RFSMS Scheme. 
VCK would like to acknowledge financial assistance from UGC, New Delhi through a major research project 
and also Associateship of IUCAA, Pune, India.

\end{document}